# Simultaneous generation of two spin-wave-photon entangled states in an atomic ensemble

*Yuelong Wu, Long Tian, Zhongxiao Xu, Wei Ge, Lirong Chen, Shujing Li, Haoxiang Yuan, Yafei Wen, Hai Wang\*, Changde Xie, Kunchi Peng*

*The State Key Laboratory of Quantum Optics and Quantum Optics Devices, Institute of Opto-Electronics, Shanxi University, Taiyuan, 030006, People's Republic of China*

The generation and storage of entangled photons play important roles in quantum information technique. Spontaneous Raman scattering (SRS) in atomic ensembles provides a promising method to generate entangled photons capable of storage. In the past experiments, a spin-wave-photon entangled state is produced via SRS in an atomic ensemble, with which a pair of entangled photons is obtained. Here, we report a scheme of simultaneously generating two spin-wave-photon entangled states in an atomic ensemble by collecting Stokes photons at two different directions. Based on the obtained two atom-photon entangled sources, we generate a three-photon GHZ polarization-entangled state and conditionally prepare a polarization-entangled photon pair, respectively.

The entangled photon pairs are the crucial resources in linear optical quantum computations (LOQC) and quantum communications (QC) [1-5]. However, the probabilistic generations of entangled photons limit their applications in the real world [1, 6-7]. For solving this problem, a promising scheme is to effectively store the entangled photons in atomic or solid-state ensembles for a desired time [1-5].

Spontaneous Raman scattering (SRS) in an atomic ensemble can emit a single photon and simultaneously create a single spin-wave excitation [8-20]. The emitted photon can be directly stored in atomic ensembles via EIT with a larger storage efficiency up to 20-50% [11-13]. The correlation between the emitted photons and the spin-wave excitations forms the physical fundament of generating the spin-wave-photon (atom-photon) entanglement [3]. In past experiments [13-17], an atom-photon entangled state has been generated from a cold atomic ensemble, with which a pair of entangled photons is obtained. However, in some quantum information protocols, the three-qubit GHZ state [18] and on-demand entanglement source [19] are required, whose preparation relies on the simultaneous generations of two or more entangled photon pairs capable of storages.

On the other hand, although the quantum repeater node [17] has been experimentally demonstrated by using two atom-photon entangled states simultaneously generated from two cold atomic ensembles, the storage lifetime of the entanglement is very short (~ 6 μs), which is not enough for the long-distance quantum communication [17]. For solving this problem, we can firstly produce a pair of polarization-entangled photons in a heralded or conditional manner, and then mapped the entanglement into atomic-ensemble memories via the long-lived (1-ms) EIT quantum storage scheme [20].

In this letter, we present an experimental demonstration of the simultaneous generation of two atom-photon entangled states in a cold atomic ensemble via SRS induced by a write laser pulse. In contrast to past experiments [13, 16-17], which achieved the high-fidelity atom-photon entanglement by encoding the photonic qubit in two spatial modes of a single photon, the presented experiment employs two photonic polarization states to encode a photonic qubit. Thus, the experimental complex is significantly reduced and the two atom-photon entangled states are obtained by collecting Stokes photons at two different directions with a small angle relative to quantum axis. The measured Bell parameters for the two entangled states are $2.77 \pm 0.01$ and $2.64 \pm 0.01$, respectively. Based on the two high-fidelity entangled states, we generate a three-photon GHZ polarization-entangled state and conditionally prepare a polarization-entangled photon pair, respectively.

The experimental setup and relevant atomic levels for generating two independent atom-photon entangled states, which are the fundamental entanglement sources in the experiment, are shown in Fig.1 (a-c). The atomic ensemble is a cloud of cold $^{87}$Rb atoms loaded by a magneto-optical trap (MOT). After loading the atoms into the MOT and subsequent cooling, a magnetic field $B \approx 200 mG$ is applied along z-direction to define the quantum axis. The cold atoms are prepared in the initial state $|a\rangle$ and then a z-polarized writing laser pulse with $10 MHz$ red-detuned to $|a\rangle \rightarrow |e_2\rangle$ transition is applied onto the atoms. Without considering Zeeman sublevels (see Fig.1 (b)), the write pulse will induce a spontaneous Raman transition $|a\rangle \rightarrow |b\rangle$ via $|e_2\rangle$, in which Stokes photons are emitted and collective (spin-wave) excitations are created at the same time. If a single Stokes photon is collected and detected in a spatial mode $S_i$ at a small angle ($\pm 0.4°$ for $i=1,2$ in the presented experiment) relative to z axis, a single spin-wave (SW) excitation correlated with the single photon will be generated in the spatial mode $A_i$. The wave-vector of the SW mode $A_i$ is $\vec{k}_{A_i} = \vec{k}_w - \vec{k}_{S_i}$, where, $\vec{k}_w$ and $\vec{k}_{S_i}$ are the wavevectors of write field and Stokes field $S_i$, respectively.

If the Zeeman sublevels (Fig.1(c)) are considered, the entanglement between the single SW excitation and the single Stokes photon will be observed. Assuming that the atoms are initially prepared in $|a, m_a\rangle$ state ($m_a = 0, \pm 1$), two spontaneous Raman transitions will occur: one is $|a, m_a\rangle \rightarrow |b, m_b = m_a - 1\rangle$ via $|e_2, m_a\rangle$, which generates a correlated pair of a single $\sigma^+$-polarized Stokes photon in $S_i$ mode and a SW excitation $|\psi^{m_a, m_b = m_a - 1}\rangle$ in $A_i$ mode and another one is $|a, m_a\rangle \rightarrow |b, m_b = m_a + 1\rangle$ via $|e_2, m_a\rangle$, which generates a correlated pair of a single $\sigma^-$-polarized Stokes

photon in $S_i$ mode and one SW excitation $|\psi^{m_a,m_b=m_a+1}\rangle$ in $A_i$ mode. The SW is represented by:

$$|\psi^{m_a,m_b}\rangle_{A_i} = \sqrt{\frac{2F_a+1}{N}} \sum_{j=1}^{N_m} |a_1^{m_a}\rangle_{A_i} \cdots |b_j^{m_b=m_a+\alpha}\rangle_{A_i} \cdots |a_{N_{m_a}}^{m_a}\rangle_{A_i} e^{-\vec{k}_{A_i}\cdot\vec{r}_j} \quad (1)$$

which is associated with the coherence $|a,m_a\rangle \leftrightarrow |b,m_b=m_a+\alpha\rangle$, where $N$ is the atomic number in the spatial mode $A_i$, $N_{m_a}=\frac{N}{2F_a+1}$, $F_a=1$. Considering the general case that the atoms are prepared in $|a\rangle$ with equal probability in the three Zeeman sublevels $|m_a=-1\rangle$, $|m_a=0\rangle$ and $|m_a=1\rangle$, the emission of a single $\sigma^+$-polarized Stokes photon will create incoherent superposition of SWs $|\psi^{-1,-2}\rangle_{A_i}$, $|\psi^{0,-1}\rangle_{A_i}$ and $|\psi^{1,0}\rangle_{A_i}$, while, the emission of $\sigma^-$-polarized Stokes photon will create incoherent superposition of SWs $|\psi^{-1,0}\rangle_A$ $|\psi^{0,1}\rangle_A$ and $|\psi^{1,2}\rangle_A$. If accounting for the Celebsch-Gorden efficiencies of the relative transitions and omitting the SWs ($|\psi^{-1,-2}\rangle_{A_i}$ and $|\psi^{1,2}\rangle_A$) not retrieved in the read process, the two incoherent superpositions can be written as (see Supplemental Material [21]): $|\psi^+\rangle_{A_i} = \sqrt{\frac{3}{7}}|\psi^{-1,0}\rangle_{A_i} + \sqrt{\frac{4}{7}}|\psi^{0,1}\rangle_{A_i}$ and $|\psi^-\rangle_{A_i} = \sqrt{\frac{4}{7}}|\psi^{0,-1}\rangle_{A_i} + \sqrt{\frac{3}{7}}|\psi^{1,0}\rangle_{A_i}$, respectively. The two SWs $|\psi^\pm\rangle_{A_i}$ are orthogonal and thus can be used to encode the memory qubit. Under the condition of the excitation probability $\chi \ll 1$, the polarization state of the single photon $S_i$ will be entangled with the memory qubit in $A_i$ mode. Neglecting the vacuum state and high order excitations, the entanglement state between the memory qubit and photonic polarization state can be written as

6[21]:

$$|\Phi\rangle_{S_i-A_i} = \frac{1}{\sqrt{2}}\left(|L\rangle_{S_i}|\psi^+\rangle_{A_i} + |R\rangle_{S_i}|\psi^-\rangle_{A_i}\right) \qquad (2)$$

where $|R\rangle/|L\rangle$ represents $\sigma^+-/\sigma^--$ polarized single photons. After a storage time $\tau$, a strong read laser pulse resonating with atomic $|b\rangle \to |e_1\rangle$ transition is applied and thus the single SW excitation $|\psi^+\rangle_{A_i}$ ($|\psi^-\rangle_{A_i}$) is transferred into single $\sigma^+$ ($\sigma^-$) polarized anti-Stokes photon in $S_i{'}$ mode. The $S_i{'}$ mode has the wave-vector of $\vec{k}_i{'} = -\vec{k}_i$ under the condition of the counterpropagation of write and read beams (see Fig.1 (a)). We place a $\lambda/4$-plate in the path of $S_1$, $S_2$, $S_1{'}$ and $S_2{'}$ mode, respectively, to transform their circular polarizations into linear polarizations ($S_1$ and $S_2$: $R$ ($L$) $\to$ $H$ ($V$); $S_1{'}$and $S_2{'}$: $R$ ($L$) $\to$ $V$ ($H$)), where $H/V$ represents horizontal/vertically polarization. The entanglement state between the Stokes and anti-Stokes photons can be written as:

$$|\Psi(\tau)\rangle_{S_i{'}-S_i} = \frac{1}{\sqrt{2}}\left(|H\rangle_{S_i{'}}|H\rangle_{S_i} + |V\rangle_{S_i{'}}|V\rangle_{S_i}\right). \qquad (3)$$

The quality of the entanglement can be judged by the Bell-CHSH inequality [28]: $S_i^{CHSH} = |E(\theta_{s_i},\theta_{as_i}) - E(\theta_{s_i},\theta_{as_i}{'}) + E(\theta_{s_i}{'},\theta_{as_i}) + E(\theta_{s_i}{'},\theta_{as_i}{'})| < 2$, where $E(\theta_{s_i},\theta_{as_i})$ is the correlation functions between the anti-Stokes and Stokes photons [21], $\theta_{s_i}$ and $\theta_{as_i}$ are the polarization-angle settings of the Stokes photon $S_i$ and anti-Stokes photon $S_i{'}$, respectively. For suppressing the double or multiple emissions, we chose a low excitation probability ($\chi \approx 0.014$) to perform the measurements. The measurements are carried out in a cyclic

fashion [21]. In the case of the canonical settings $\theta_{s_i} = 0^0$, $\theta_{as_i} = 22.5^0$, $\theta_{s_i}' = 45^0$ and $\theta_{as_i}' = 67.5^0$, we measure $E(\theta_{s_i}, \theta_{as_i})$ and then obtain the Bell parameters, which are $S_1^{CHSH} = 2.77 \pm 0.01$ ($2.75 \pm 0.01$) and $S_2^{CHSH} = 2.64 \pm 0.01$ ($2.67 \pm 0.01$) for the storage time of $\tau = 30ns$ ($230ns$), respectively, both of them significantly violate the Bell-CHSH inequality $|S^{CHSH}| < 2$. The measured generation rates for the entangled photon pairs $S_1/S_1'$ and $S_2/S_2'$ are ~60/s (~48/s) at $\tau = 30ns$ ($230ns$).

As shown in Fig.1(d), the two Stokes photons $S_1$ and $S_2$, coming from the atom-photon entanglement sources $|\Psi(\tau)\rangle_{A_1-S_1}$ and $|\Psi(\tau)\rangle_{A_2-S_2}$, pass through two single-mode fibers, respectively, and then are overlapped on a polarization beam splitter (PBS) of transmitting $H$ and reflecting $V$ polarization to perform a two-photon interference. If the two Stokes photons are in the same polarization $H$ or $V$, they will exit from two different output ports of PBS. Thus, a four-particle entangled state will be formed [22], which can be written as:

$$|\Psi\rangle_{G4} = \frac{1}{\sqrt{2}}\left(|\psi^+\rangle_{A_1}|\psi^+\rangle_{A_2}|H\rangle_1|H\rangle_2 + |\psi^-\rangle_{A_1}|\psi^-\rangle_{A_2}|V\rangle_1|V\rangle_2\right) \quad (4)$$

where, the subscript 1 and 2 denote the emitting photons from the two output ports.

For obtaining the three-photon GHZ entanglement, we may perform a suitable polarization measurement on the photon 1. As shown in Fig.1 (d), the polarization of the photon 1 is rotated by $45^0$ with the $\lambda/2$ plate at the front of PBS1. In this case, if $D_{H1}$ registers a photon, the photon 1 is projected into the polarization state $|H'\rangle = (|H\rangle + |V\rangle)/\sqrt{2}$ and at the same time, the four-particle entangled state is projected into a tri-particle GHZ entangled state [21]:

$$|\Psi\rangle_{G3} = \frac{1}{\sqrt{2}}\left(|\psi^+\rangle_{A_1}|\psi^+\rangle_{A_2}|H\rangle_2 + |\psi^-\rangle_{A_1}|\psi^-\rangle_{A_2}|V\rangle_2\right) \quad (5)$$

After a storage time $\tau$, the SWs $|\psi^+\rangle_{A_1}$ and $|\psi^-\rangle_{A_1}$ ($|\psi^+\rangle_{A_2}$ and $|\psi^-\rangle_{A_2}$) are retrieved into photonic polarization states $|H\rangle_3$ and $|V\rangle_3$ ($|H\rangle_4$ and $|V\rangle_4$) by dynamic EIT and the tri-particle GHZ state is transferred into a three-photon GHZ entangled state

$$|\Psi\rangle_{G3} = \frac{1}{\sqrt{2}}\left(|H\rangle_2|H\rangle_3|H\rangle_4 + |V\rangle_2|V\rangle_3|V\rangle_4\right) \quad (6)$$

To evaluate the quality of the entangled state, we apply the two-setting witness [23]:

$$W_{GHZ3} = \frac{3}{2}I^{\otimes 3} - \sigma_x^{(2)}\sigma_x^{(3)}\sigma_x^{(4)} - \frac{1}{2}\left(\sigma_z^{(2)}\sigma_z^{(3)} + \sigma_z^{(3)}\sigma_z^{(4)} + \sigma_z^{(2)}\sigma_z^{(4)}\right) \quad (7)$$

where, $I$ is a 2-dimensional identity matrix, $\sigma_x^{(2)}\sigma_x^{(3)}\sigma_x^{(4)}$ represents a joint measurement of linear polarization $H'/V'$ on photons 2, 3 and 4, $\sigma_z^{(2)}\sigma_z^{(3)}$ ($\sigma_z^{(3)}\sigma_z^{(4)}$, $\sigma_z^{(2)}\sigma_z^{(4)}$) denotes a joint measurement of linear polarization $H/V$ on photons 2 and 3 (3 and 4, 2 and 4), where, $|V'\rangle = (|H\rangle - |V\rangle)/\sqrt{2}$). If the measured witness is a negative value, the tri-particle will be in GHZ entangled state [25]. The measured expectation values of the observables $E(\sigma_x^{(2)}\sigma_x^{(3)}\sigma_x^{(4)})$, $E(\sigma_z^{(2)},\sigma_z^{(3)})$, $E(\sigma_z^{(3)},\sigma_z^{(4)})$, $E(\sigma_z^{(2)},\sigma_z^{(4)})$ conditioned on detecting a photon at detector $D_{H1}$ are listed in Table 1, each expectation value is deduced from 160 to 190 fourfold-coincidence events. Note that the detecting time of the photons 1 and 2 is earlier than that of the photons 3 and 4 by a storage time $\tau = 30ns$. Substituting these data into Eq. (6), we obtain $\langle W_{GHZ3} \rangle = -0.68 \pm 0.10$, which confirms that the three photons are in the genuine GHZ entanglement state.

For demonstrating the confliction between local realism and quantum mechanics for the GHZ entanglement, we use Mermin inequality of the tri-particle [24]:

$$S_{Me} = \left| E(\sigma_y^{(2)}, \sigma_y^{(3)}, \sigma_x^{(4)}) + E(\sigma_y^{(2)}, \sigma_x^{(3)}, \sigma_y^{(4)}) + E(\sigma_x^{(2)}, \sigma_y^{(3)}, \sigma_y^{(4)}) - E(\sigma_x^{(2)}, \sigma_x^{(3)}, \sigma_x^{(4)}) \right| \leq 2 \quad (8)$$

where, for example, $E(\sigma_y^{(2)}, \sigma_y^{(3)}, \sigma_x^{(4)})$ is the expectation value of the measurement setting $\sigma_y^{(2)}\sigma_y^{(3)}\sigma_x^{(4)}$, which represents the joint measurement of circular polarization $R/L$ on photons 1 and 2 and linear polarization $H'/V'$ on photon 3, $|R\rangle = (|H\rangle + i|V\rangle)/\sqrt{2}$ and $|L\rangle = (|H\rangle - i|V\rangle)/\sqrt{2}$). The measured expectation values $E(\sigma_y^{(2)}, \sigma_y^{(3)}, \sigma_x^{(4)})$, $E(\sigma_y^{(2)}, \sigma_x^{(3)}, \sigma_y^{(4)})$, $E(\sigma_x^{(2)}, \sigma_y^{(3)}, \sigma_y^{(4)})$ and $E(\sigma_x^{(2)}, \sigma_x^{(3)}, \sigma_x^{(4)})$ conditioned on detecting a photon at detector $D_{H1}$ are shown in Table 1, each value is also deduced from 160 to 190 fourfold-coincidence events. From the measured data, we obtain the Mermin parameter of $S_{Me} = 3.14 \pm 0.12$, which violates the limit of the local realism by 9 standard deviations. Because we use Pauli measurements in the Mermin inequality, this violation also confirms that three photons are in a genuine GHZ entangled state. The conclusion is correct for the imperfect $\sigma_x$ and $\sigma_y$ measurements, since the measured Mermin parameter $S_{Me}$ is larger than the limit of device-independent witness of genuine tri-particle entanglement of $2\sqrt{2}$ [25].

We now describe the conditional preparation of the two-photon polarized-entangled state via Bell state measurement (BSM). As shown in Fig.1 (d), two $\lambda/2$ plates are inserted in the paths of modes 1 and 2 to rotate the polarizations of photons 1 and 2 by $45^0$, respectively. In this case, the measurement of linear polarization $H'/V'$ on photon 1 (2) will be implemented by detectors $D_{H1}/D_{V1}$ ($D_{H2}/D_{V2}$). Conditioned on detecting a coincidence at detectors $D_{H1}D_{H2}$ or $D_{V1}D_{V2}$, the photons 1 and 2 are projected into the Bell state $\Phi_{12}^+ = (|H'\rangle_1|H'\rangle_2 + |V'\rangle_1|V'\rangle_2)/\sqrt{2}$ and thus the two stored SW qubits are projected into the entangled state [21]:

$$|\Phi^+\rangle_{A_1A_2} = \frac{1}{\sqrt{2}}\left(|\psi^+\rangle_{A_1}|\psi^+\rangle_{A_2} + |\psi^-\rangle_{A_1}|\psi^-\rangle_{A_2}\right). \tag{9}$$

After a storage time $\tau$, a read laser pulse is applied to transfer the state $|\Phi^+\rangle_{A_1A_2}$ into the two-photon polarization-entangled state:

$$|\Psi_{34}\rangle = \frac{1}{\sqrt{2}}\left(|H\rangle_3|H\rangle_4 + |V\rangle_3|V\rangle_4\right). \tag{10}$$

Actually, there are double excitations in either the spin-wave mode $A_1$ or $A_2$ [21], which will induce error events in the BSM and then the success probability of the conditionally prepared entangled state $|\Phi^+\rangle_{A_1A_2}$ is reduce to 1/2 [5, 27, 22]. The error events can be eliminated at the stage of entanglement verification of the state $|\Phi^+\rangle_{A_1A_2}$ by the fourfold coincident of photons 1, 2, 3 and 4 [17].

We measure correlation functions $E(\theta_3, \theta_4)$ and then obtain the Bell parameter $S^{CHSH} = |E(\theta_3, \theta_4) - E(\theta_3', \theta_4) + E(\theta_3, \theta_4') + E(\theta_3', \theta_4')|$ for the entangled state $|\Psi_{34}\rangle$ at different storage times $\tau$, where $\theta_3$ and $\theta_4$ are the

polarization angles of the photon 3 and 4, respectively. The measured results are shown in Table 2. At $\tau = 30 ns$, we obtain a maximal value of $S^{CHSH} = 2.40 \pm 0.07$, which violates Bell-CHSH inequality by ~5.7 standard deviations. The measured rate for the pair of entangled photons 3 and 4 conditioned on detecting the coincidence event in detectors $D_{H1}D_{H2}$ or $D_{V1}D_{V2}$ is $R_{3,4} \approx 22$/hour for $\tau = 30 ns$, while the measured rate of the coincidence between two detectors $D_{H1}D_{H2}$ or $D_{V1}D_{V2}$ is $C_{1,2} \approx 1.78 \times 10^4$ /hour. So, the success probability of conditionally preparing entangled photons 3 and 4 is $r_{34} = R_{3,4} / C_{1,2} \approx 1.24 \times 10^{-3}$, which is far lower than the theoretically calculated limit of 1/2. The difference between the theoretical expectation and the experimental value results from the imperfect retrieval efficiencies of the two SWs and detection efficiencies of the detectors $D_3$ ($D_{H3}$/ $D_{V3}$) and $D_4$ ($D_{H4}$/ $D_{V4}$).

In conclusion, we simultaneously produced two atom-photon entangled states in a $^{87}$Rb cold atomic ensemble via SRS. Based on the entangled states, we generate the three-photon polarization entangled state and conditionally prepare the polarization-entangled photon pair, respectively. Especially, the polarization-entangled photon pair can be generated after a controllable storage time, which is very useful for some quantum information processing [26]. Although the measured success probability of the conditionally prepared entangled pair is lower in the presented work, it can be increased by improving the retrieval and detection efficiencies. In the experiment, we produce only two spin-wave-photon entanglement sources by collecting the Stokes photons at two directions. If the Stokes photons are collected at more directions, more than two pairs of entangled and/or correlated photons can be created. Relying on these multiple photon pairs and using EIT-based quantum storages, it is possible to obtain large-size cluster states or on-demanded entanglement sources, which are the necessary quantum resources for realizing some complex quantum information protocols [24-25].

Acknowledgement: We acknowledge funding support from the 973 Program (2010CB923103), the National Natural Science Foundation of China (No. 11475109, 11274211, 60821004).

*Corresponding author: wanghai@sxu.edu.cn

**References**

1. J.-W. Pan, Z.-B. Chen, C.-Y. Lu, H. Weinfurter, A. Zeilinger, and M. Żukowski, Rev. Mod. Phys. **84,** 777 (2012).

2. P. Kok, W. J. Munro, K. Nemoto, T. C. Ralph, J. P. Dowling and G. J. Milburn, Rev. Mod. Phys. **79,** 135 (2007).

3. N. Sangouard, C. Simon, H. de Riedmatten, and N. Gisin, Rev. Mod. Phys. **83,** 33 (2011).

4. N. Sangouard, C. Simon, B. Zhao, Y.-A. Chen, H. de Riedmatten, J.-W. Pan, and N. Gisin, Phys. Rev. A. **77,** 062301 (2008).

5. Z.-B. Chen, B. Zhao, Y.-A. Chen, J. Schmiedmayer, and J.-W. Pan, Phys. Rev. A **76,** 022329 (2007).

6. C. Wagenknecht, C.-M. Li, A. Reingruber, X.-H. Bao, A. Goebel, Y.-A. Chen, Q. Zhang, K. Chen and J.-W. Pan, Nature Photon. **4,** 549 (2010).

7. S. Barz, G. Cronenberg, A. Zeilinger and P. Walther, Nature Photon. **4,** 553–556 (2010).

8. L.-M. Duan, M. Lukin, J. I. Cirac and P. Zoller, Nature **414,** 413 (2001).

9. A. Kuzmich, W. P. Bowen, A. D. Boozer, A. Boca, C. W. Chou, L.-M. Duan and H. J. Kimble, Nature **423,** 731 (2003).

10. J. Laurat, H. de Riedmatten, D. Felinto, C.-W. Chou, E. W. Schomburg and H. J. Kimble, Opt. Express **14,** 006912 (2006).

11. T. Chanelière, D. N. Matsukevich, S. D. Jenkins, S.-Y. Lan, T. A. B. Kennedy and A. Kuzmich, Nature **438,** 833 (2005).

12. S.-Y. Zhou, S.-C. Zhang, C. Liu, J. F. Chen, J.-M. Wen, M. M. T. Loy, G.K. L. Wong and S.-W. Du, Opt. Express **20,** 24142 (2012).

13. D.-S. Ding, W. Zhang, Z.-Y. Zhou, S. Shi, B.-S. Shi and G.-C. Guo, Nature Photon. **10,** 1038 (2015).

14. D. N. Matsukevich, T. Chanelière, M. Bhattacharya, S.-Y. Lan, S. D. Jenkins, T. A. B. Kennedy and A. Kuzmich, Phys. Rev. Lett. **95,** 040405 (2005).

15. H. de Riedmatten, J. Laurat, C. W. Chou, E. W. Schomburg, D. Felinto and H. J. Kimble, Phys. Rev. Lett. **97,** 113603 (2006).

16. S. Chen, Y.-A Chen, B. Zhao, Z.-S. Yuan, J. Schmiedmayer and J.-W. Pan, Phys. Rev. Lett. **99,** 180505 (2007).

17. Z.-S. Yuan, Y.-A. Chen, B. Zhao, S. Chen, J. Schmiedmayer and J.-W. Pan, Nature **454,**

1098 (2008).

18. D. E. Browne and T. Rudolph, Phys. Rev. Lett. **95,** 010501 (2005).

19. J. Minář, H. de Riedmatten and N. Sangouard, Phys. Rev. A **85,** 032313 (2012).

20. Z.-X. Xu, Y.-L. Wu, L. Tian, L.-R. Chen, Z.-Y. Zhang, Z.-H. Yan, S.-J. Li, H. Wang, C.-D. Xie and K.-C. Peng, Phys. Rev. Lett. **111,** 240503 (2013).

21. See Supplemental Material at [URL will be inserted by publisher] for experimental details, theoretical discussions of atom-photon entanglement states, generation of the three-photon polarization entangled state, conditional preparation of the polarization-entangled photon pair and the success probability of the preparation.

22. Z. Zhao, T. Yang, Y.-A. Chen, A.-N. Zhang, M. Żukowski and J.-W. Pan, Phys. Rev. Lett. **91,** 180401 (2003).

23. G. Tóth, and O. Gühne, Phys. Rev. A **72,** 022340 (2005).

24. N. D. Mermin, Phys. Rev. Lett. **65,** 1838 (1990).

25. J. D. Bancal, N. Gisin, Y.-C. Liang and S. Pironio, Phys. Rev. Lett. **106,** 250404 (2011).

26. A. I. Lvovsky, B. C. Sanders, and W. Tittel, Nature Photon. **3,** 706 (2009).

27. B. Zhao, Z.-B. Chen, Y.-A. Chen, J. Schmiedmayer and J.-W. Pan, Phys. Rev. Lett. **98,** 240502 (2007).

28. J.-W. Pan, C. Simon, Č. Brukner and A. Zeilinger, Nature **410,** 1067 (2001).

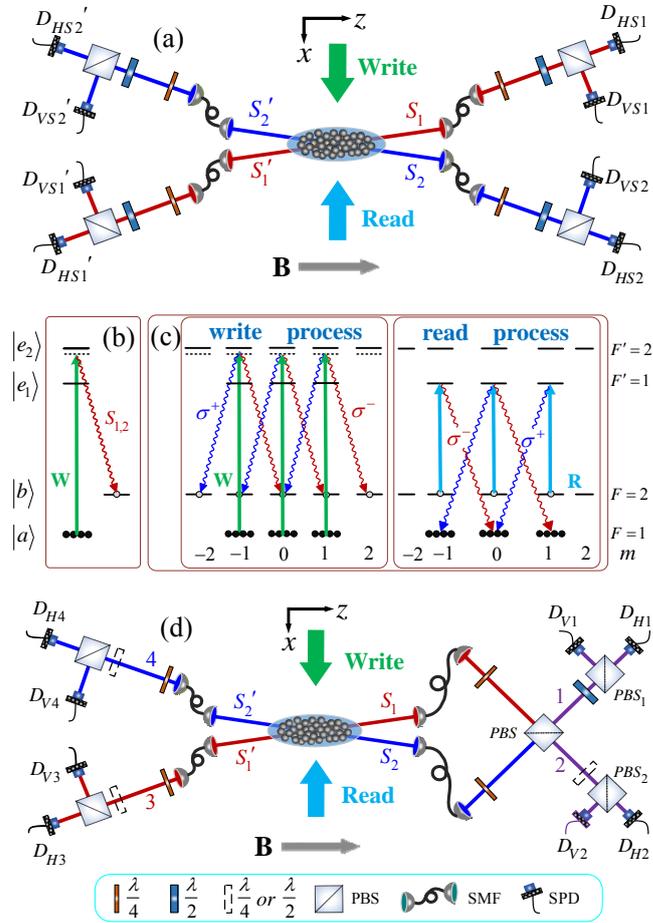

**Fig.1 Overview of the experiment.** (a) Experimental setup of two atom-photon entangled states. (b) and (c) Relevant atomic levels without and with considering Zeeman sublevels. (d) Experimental setup for either generating three-photon GHZ entanglement or conditionally preparing polarization-entangled photon pairs. PBS: polarization beam splitter; SMF: single mode fiber; SPD: single photon detector.

**Table1 Measured witness and Mermin correlations for $\tau = 30 ns$. The errors represent $\pm 1$ standard deviation.**

| | | | |
|---|---|---|---|
| $E\left(\sigma_z^{(2)}, \sigma_z^{(3)}\right)$ | $0.92 \pm 0.02$ | $E\left(\sigma_x^{(2)}, \sigma_y^{(3)}, \sigma_y^{(4)}\right)$ | $-0.80 \pm 0.06$ |
| $E\left(\sigma_z^{(3)}, \sigma_z^{(4)}\right)$ | $0.89 \pm 0.03$ | $E\left(\sigma_y^{(2)}, \sigma_x^{(3)}, \sigma_y^{(4)}\right)$ | $-0.77 \pm 0.06$ |
| $E\left(\sigma_z^{(2)}, \sigma_z^{(4)}\right)$ | $0.94 \pm 0.02$ | $E\left(\sigma_y^{(2)}, \sigma_y^{(3)}, \sigma_x^{(4)}\right)$ | $-0.77 \pm 0.06$ |
| $E\left(\sigma_x^{(2)}, \sigma_x^{(3)}, \sigma_x^{(4)}\right)$ | $0.80 \pm 0.06$ | $E\left(\sigma_x^{(2)}, \sigma_x^{(3)}, \sigma_x^{(4)}\right)$ | $0.80 \pm 0.06$ |

**Table2 Measured correlation function $E(\theta_3,\theta_4)$ for the conditionally prepared entangled state.** The errors represent ±1 standard deviation.

|  | 30$ns$ | 230$ns$ | 430$ns$ |
|---|---|---|---|
| $E(0°,22.5°)$ | 0.55±0.04 | 0.55±0.04 | 0.63±0.06 |
| $E(0°,67.5°)$ | −0.66±0.05 | −0.67±0.05 | −0.59±0.05 |
| $E(45°,22.5°)$ | 0.57±0.04 | 0.44±0.04 | 0.26±0.02 |
| $E(45°,67.5°)$ | 0.63±0.04 | 0.61±0.05 | 0.57±0.05 |
| $S^{CHSH}$ | 2.40±0.07 | 2.27±0.08 | 2.05±0.08 |

# Supplemental Material

# Simultaneous generation of two spin-wave-photon entangled states in an atomic ensemble

Yuelong Wu, Long Tian, Zhongxiao Xu, Wei Ge, Lirong Chen, Shujing Li, Haoxiang Yuan, Yafei Wen, Hai Wang*, Changde Xie, Kunchi Peng

The State Key Laboratory of Quantum Optics and Quantum Optics Devices, Institute of Opto-Electronics, Shanxi University, Taiyuan, 030006, People's Republic of China

## I. Experimental details

As shown in Fig.1 (a) of the main text, $\sim 10^9$ $^{87}$Rb atoms are trapped in a two-dimension magneto-optical trap (MOT) with a size of $\sim 5\times 2\times 2 mm^3$, a temperature of ~130μK and an optical density of about 7. The write and read beams counterpropagate through the atoms, whose diameters (powers) in the MOT are ~3mm (1mW) and 3.3mm (~50 mW), respectively. The atoms are optically pumped into the initial level $|a\rangle$ by $\sigma^{\pm}$-polarized laser beams $P_1$ and $P_2$ (not shown in Fig.1 (a)), which are overlapped at a PBS and then collinearly go through the atoms at an angle of 2° to the z-axis. The frequencies of $P_1$ and $P_2$ are tuned on the $|b\rangle \leftrightarrow |e_2\rangle$ and $|b\rangle \leftrightarrow |e_1\rangle$ transitions, respectively, and the power of both $P_1$ and $P_2$ is kept at ~60mW.

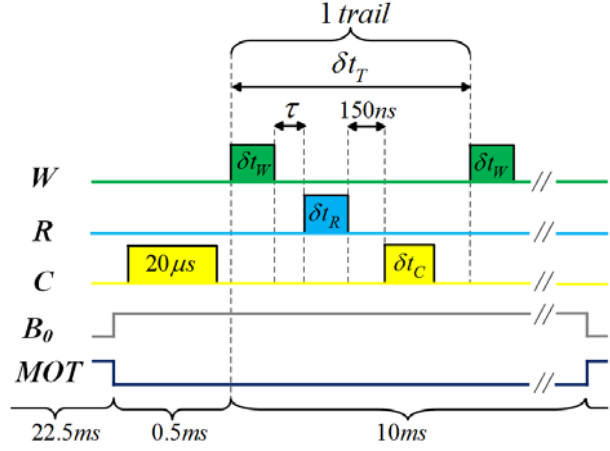

**Fig.S1 The time sequence of an experimental cycle.**

The time sequence of the experimental procedure is shown in Fig.S1. The experimental cycle repeats with a frequency of 30 Hz. Each experimental cycle contains a 23-ms preparation stage and a 10-ms experiment run. During the preparation stage, the atomic ensemble are trapped in the MOT for 22 ms and further cooled by a Sisyphus cooling for 0.5ms. Then the MOT (beams and magnetic field) is turned off and the guiding magnetic field *B=200mG* is applied. At the end of this stage, $P_1$ and $P_2$ laser beams are switched on for 20 μs to pump the atoms into the level $|a\rangle$. After the preparation stage, the 10-ms experiment run which contains *n* trails starts. In each trail, a write laser pulse with a length of $\delta t_W = 70ns$ is applied to generate correlated pairs of a single Stokes photon and a single SW excitation. After a storage time $\tau$, a read laser pulse with a length of $\delta t_R = 100 ns$ is applied to retrieve the stored SWs. At the end of each trail, the atoms are pumped back to the $|a\rangle$ by switching on the laser beams $P_1$ and $P_2$ for a duration time of $\delta t_c = 200 ns$. For the case of $\tau = 30 ns$ ($\tau = 230, 430 ns$), the 10-ms experimental run contains $n$=10000 ($n$=8333) trails, thus 1-s experimental sequence contains $N$=300000 ($N$=24990) trails.

The overall detection efficiency $\eta_{S_1}$, $\eta_{S_2}$, $\eta_{S_1'}$ and $\eta_{S_2'}$ for detecting the photons in $S_1$, $S_2$, $S_1'$ and $S_2'$ modes is ~30%, which includes the transmissions of the optical filters (80%), the coupling efficiency of fiber coupler (80%), the efficiency of fiber coupling to the single-photon detectors (95%) and the quantum efficiency of the single-photon detectors (50%). The retrieval efficiency of both SWs is $R_i \approx 20\%$ for the storage time of $\tau = 30 ns$. For the excitation probability of $\chi \approx 0.014$, the calculated pair rate of the entangled state $\Phi_{S_i-S_i'}$ is $r_{S_i S_i'} = \eta_{S_i} \eta_{S_i'} \chi R_i N \approx 76/s$, which is reasonably consistent with the measured data of 60/s. The success probability of conditionally preparing entangled photons 3 and 4 can be evaluated by $r_{34} = 0.5 \eta_{D3} \eta_{D4} R_3 R_4 \approx 1.8 \times 10^{-3}$, which also is basically consistent with the experimental value of $r_{34} \approx 1.24 \times 10^{-3}$.

## II. Atom-photon entanglement state

When atoms are initially pumped in the state $|a^{m_a}\rangle$ ($m_a = \pm 1, 0$ is the magnetic quantum number of the level $|a\rangle$, $|a^{m_a}\rangle$ denotes the state $|a, m_a\rangle$), a write pulse will create an atom-photon joint state $\rho_{S_i-A_i} = |0\rangle\langle 0| + |\Phi\rangle_{S_i-A_i}\langle\Phi|$ via

SRS. In the joint state, the nonvacuum part is an entangled state:

$$|\Phi\rangle_{S_i-A_i} = \chi\left(\cos\eta \hat{S}_+^\dagger \hat{a}_L^\dagger + \sin\eta \hat{S}_-^\dagger \hat{a}_R^\dagger\right)|0\rangle_{S_i}^A |0\rangle_{A_i}^P + o(\chi^2) \quad (S1)$$

where $S_i$ and $A_i$ ($i=1,2$) are the spatial modes of Stokes fields and spin-waves, whose directions are described by the wave-vectors $\vec{k}_{S_i}$ and $\vec{k}_{A_i}$, respectively; $|0\rangle_{A_i}^P$ and $|0\rangle_{S_i}^A = \otimes^i |a\rangle_{S_i}^i = \otimes^i |a^{m=-1}\rangle_{S_i}^i \otimes^i |a^{m=0}\rangle_{S_i}^i \otimes^i |a^{m=1}\rangle_{S_i}^i$ denote the vacuum states of the spin-wave mode $A_i$ and Stokes light mode $S_i$, respectively. The parameter is obtained from $\cos\eta = \frac{\sqrt{X_{\alpha=-1}^2}}{\sqrt{X_{\alpha=-1}^2 + X_{\alpha=+1}^2}}$ [SR1], where

$X_{\alpha=\pm 1}^2 = X_{\alpha=\pm 1}^2(m=1) + X_{\alpha=\pm 1}^2(m=0) + X_{\alpha=\pm 1}^2(m=-1)$, $\alpha = \pm 1$ is photon helicity, $X_\alpha(m) = C_{m,0,m}^{F_a,1,F_{e_2}} C_{m,\alpha,m+\alpha}^{F_{e_2},1,F_b}$ is the product of the relevant Clebsch-Gordan coefficient for the transition. If the initial state is incoherently distributed in the various $|a^m\rangle$ states, the spin-wave operators $\hat{S}_+^\dagger$ and $\hat{S}_-^\dagger$ are given by [SR1]:

$$\hat{S}_+^\dagger = \frac{X_{+1}(-1)}{X_{+1}} S_{+1}^\dagger(-1) + \frac{X_{+1}(0)}{X_{+1}} S_{+1}^\dagger(0) + \frac{X_{+1}(1)}{X_{+1}} S_{+1}^\dagger(1) \quad (S2a)$$

$$\hat{S}_-^\dagger = \frac{X_{-1}(-1)}{X_{-1}} S_{-1}^\dagger(-1) + \frac{X_{-1}(0)}{X_{-1}} S_{-1}^\dagger(0) + \frac{X_{-1}(1)}{X_{-1}} S_{-1}^\dagger(1) \quad (S2b)$$

where, $\hat{S}_\alpha^\dagger(m) = \sqrt{\frac{2F_a+1}{N}} \sum_i |b^{\alpha+m}\rangle_i \langle a^m|$. In this case, we have:

$$\hat{S}_+^\dagger |0\rangle = |\psi^+\rangle_{A_i} = \left(X_{+1}(-1)|\psi^{-1,0}\rangle_{A_i} + X_{+1}(0)|\psi^{0,1}\rangle_{A_i} + X_{+1}(1)|\psi^{1,2}\rangle_{A_i}\right)/X_{+1} \quad (S3a)$$

$$\hat{S}_-^\dagger |0\rangle = |\psi^-\rangle_{A_i} = \left(X_{-1}(-1)|\psi^{-1,-2}\rangle_{A_i} + X_{-1}(0)|\psi^{0,-1}\rangle_{A_i} + X_{-1}(1)|\psi^{1,0}\rangle_{A_i}\right)/X_{-1} \quad (S3b)$$

where, $|\psi^+\rangle_{A_i}$ ($|\psi^-\rangle_{A_i}$) represents a mixed spin-wave state with one excitation, the

SW $|\psi^{m_a,m_b}\rangle_{A_i} = \sqrt{\frac{2F_a+1}{N}} \sum_{j=1}^{N_m} |a_1^{m_a}\rangle_{A_i} \cdots |b_j^{m_b}\rangle_{A_i} \cdots |a_{N_{m_a}}^{m_a}\rangle_{A_i} e^{-\vec{k}_{A_i}\cdot\vec{r}_j}$ associates with

the $|a^{m_a}\rangle \leftrightarrow |b^{m_b}\rangle$ coherence. Considering that the SWs $|\psi^{1,2}\rangle_{A_i}$ and $|\psi^{-1,-2}\rangle_{A_i}$ can't

be retrieved in the reading process (see Fig.1(c)), we neglect them and rewrite the spin-wave states as:

$$|\psi^+\rangle_{A_i} = \left(\frac{X_{+1}(-1)}{X_{+1}}|\psi^{-1,0}\rangle_{A_i} + \frac{X_{+1}(0)}{X_{+1}}|\psi^{0,1}\rangle_{A_i}\right) = \sqrt{\frac{3}{7}}|\psi^{-1,0}\rangle_{A_i} + \sqrt{\frac{4}{7}}|\psi^{0,1}\rangle_{A_i} \quad \text{(S4a)}$$

$$|\psi^-\rangle_{A_i} = \left(\frac{X_{-1}(0)}{X_{-1}}|\psi^{0,-1}\rangle_{A_i} + \frac{X_{-1}(1)}{X_{-1}}|\psi^{1,0}\rangle_{A_i}\right) = \sqrt{\frac{4}{7}}|\psi^{0,-1}\rangle_{A_i} + \sqrt{\frac{3}{7}}|\psi^{1,0}\rangle_{A_i} \quad \text{(S4b)}$$

So, the parameters $X_{+1}$ and $X_{-1}$ are rewritten as

$$X_{+1}^2 = X_{+1}^2(-1) + X_{+1}^2(0) \quad \text{(S5a)}$$

$$X_{-1}^2 = X_{-1}^2(0) + X_{-1}^2(1) \quad \text{(S5b)}$$

respectively, and then $\cos\eta = \sin\eta = \frac{\sqrt{2}}{2}$.

Neglecting higher order excitations, the atom-photon entanglement state $|\Phi\rangle_{S_i-A_i}$ in Eq. (1) can be rewritten as

$$|\Phi\rangle_{S_i-A_i} = \frac{1}{\sqrt{2}}\left(|\psi^+\rangle_{A_i}|L\rangle_{S_i} + |\psi^-\rangle_{A_i}|R\rangle_{S_i}\right) \quad \text{(S6)}$$

which is the maximally entangled state, where, $|R\rangle/|L\rangle$ represents $\sigma^+/\sigma^-$ polarized single photon state, which are transformed into the vertically/horizontally ($V/H$) polarized single photon by placing a $\lambda/4$-plate in the path of the mode $S_1$ ($S_2$). After a storage time $\tau$, the atom-photon entanglement state becomes

$$|\Psi(\tau)\rangle_{S_i-A_i} = \frac{1}{\sqrt{2}}\left(|H\rangle_{S_i}|\psi^+(\tau)\rangle_{A_i} + |V\rangle_{S_i}|\psi^-(\tau)\rangle_{A_i}\right) \quad \text{(S7)}$$

where,

$$|\psi_{A_i}^+(\tau)\rangle = \left(\sqrt{\frac{3}{7}}e^{-i\beta\tau}|\psi_{A_i}^{-1,0}(0)\rangle + \sqrt{\frac{4}{7}}e^{i\beta\tau}|\psi_{A_i}^{0,1}(0)\rangle\right) \quad \text{(S8a)}$$

$$\left|\psi_{A_i}^-(\tau)\right\rangle = \left(\sqrt{\frac{4}{7}} e^{-i\beta\tau}\left|\psi_{A_i}^{0,-1}(0)\right\rangle + \sqrt{\frac{3}{7}} e^{i\beta\tau}\left|\psi_{A_i}^{1,0}(0)\right\rangle\right) \quad \text{(S8b)}$$

$\beta = g\mu_B B/\hbar$ is the phase shift due to Larmor precessions of the spin waves in the magnetic field $B=200mG$. So, the atom-photon entanglement state can be rewritten as:

$$\left|\Psi(\tau)\right\rangle_{S_i-A_i} = \frac{1}{\sqrt{2}}\left(\left|H\right\rangle_{S_i}\left|\psi^+(0)\right\rangle_{A_i} + e^{-i\varphi(\tau)}\left|V\right\rangle_{S_i}\left|\psi^-(0)\right\rangle_{A_i}\right) \quad \text{(S9)}$$

where, the phase shift $\varphi(\tau) = 2\arctan\dfrac{\left(\sqrt{4/7}-\sqrt{3/7}\right)\sin\beta t}{\left(\sqrt{4/7}+\sqrt{3/7}\right)\cos\beta t}$, which is plotted in Fig.S2. In the presented experiment, we perform the polarization correlation measurements in the storage time range of $\tau = 0-500ns$. In this time range, the phase shift is very small ($\varphi(\tau) \leq 3°$), so we may neglect it in the expression of the entangled state $\left|\Psi(\tau)\right\rangle_{S_i-A_i}$.

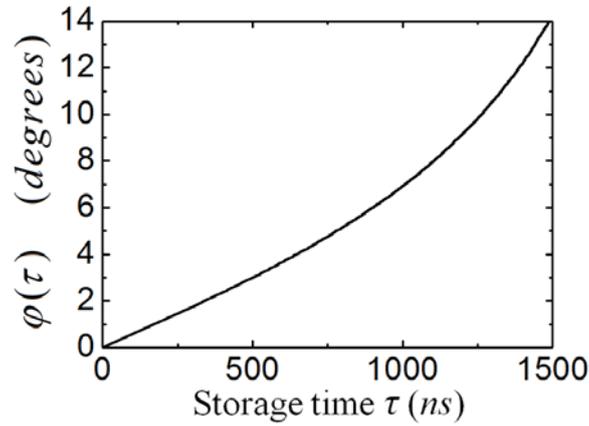

**Fig.S2 The dependence of the phase shift on the storage time $\tau$**

The single SW excitation $\left|\psi^+\right\rangle_{A_i}$ ($\left|\psi^-\right\rangle_{A_i}$) will be transferred into single $\sigma^+$ ($\sigma^-$) polarized anti-Stokes photon in the $S_i'$ field mode by applying a strong read laser pulse resonating with $|b\rangle \to |e_1\rangle$ transition onto the atoms. We also place a $\lambda/4$ -plate in the path of $S_i{'}$ mode to transform $R/L$ basis into the $H/V$ basis. After the retrieval, the obtained photon-photon entanglement state is:

$$|\Psi(\tau)\rangle_{S_i{'}-S_i} = \frac{1}{\sqrt{2}}\left(|H\rangle_{S_i{'}}|H\rangle_{S_i} + |V\rangle_{S_i{'}}|V\rangle_{S_i}\right) \tag{S10}$$

The correlation function for the entangled state $|\Psi(t)\rangle_{S_i-S_i{'}}$ is defined by:

$$E(\theta_{s_i},\theta_{as_i}) = \frac{C_{HS_iHS_i{'}}(\theta_{s_i},\theta_{as_i}) + C_{VS_iVS_i{'}}(\theta_{s_i},\theta_{as_i}) - C_{HS_iVS_i{'}}(\theta_{s_i},\theta_{as_i}) - C_{VS_iHS_i{'}}(\theta_{s_i},\theta_{as_i})}{C_{HS_iHS_i{'}}(\theta_{s_i},\theta_{as_i}) + C_{VS_iVS_i{'}}(\theta_{s_i},\theta_{as_i}) + C_{HS_iVS_i{'}}(\theta_{s_i},\theta_{as_i}) + C_{VS_iHS_i{'}}(\theta_{s_i},\theta_{as_i})} \tag{S11}$$

where, $C_{HS_iHS_i{'}}(\theta_{s_i},\theta_{as_i})$ is the number of coincidences between the detectors $D_{HS_i}$ and $D_{HS_i{'}}$ for the polarization angles $\theta_{s_i}$ and $\theta_{as_i}$ of the photons $S_i$ and $S_i{'}$.

### III. The generation of three-photon GHZ state

The four-particle entangled state $|\Psi\rangle_{G4}$ can be rewritten as:

$$|\Psi\rangle_{G4} = \frac{1}{\sqrt{2}}\left(|\Psi\rangle_{G3}^{+}|H'\rangle_1 + |\Psi\rangle_{G3}^{-}|V'\rangle_1\right) \tag{S12}$$

where, $H' = (H+V)/\sqrt{2}$ and $V' = (H-V)/\sqrt{2}$ represent $+45°$ linear polarizations, respectively, the subscript 1 denote the photon 1. If the photon 1 is found to be along the $+45°$ polarization, the two SWs and the photon 2 will be projected into the three-particle GHZ entangled state:

$$|\Psi\rangle_{G3}^{+} = \frac{1}{\sqrt{2}}\left(|\psi^{+}\rangle_{A_1}|\psi^{+}\rangle_{A_2}|H\rangle_2 + |\psi^{-}\rangle_{A_1}|\psi^{-}\rangle_{A_2}|V\rangle_2\right) \tag{S13}$$

### IV. The conditional generation of the polarization-entangled photon pair

The entanglement state $|\Psi\rangle_{G4}$ can also be rewritten as:

$$|\Psi\rangle_{G4} = \frac{1}{2}\left(|\Phi\rangle_{34}^{+}|\varphi\rangle_{12}^{+} + |\Phi\rangle_{34}^{-}|\varphi\rangle_{12}^{-}\right) \tag{S14}$$

where $|\Phi\rangle_{A_1A_2}^{\pm} = \frac{1}{\sqrt{2}}\left(|\psi^{+}\rangle_{A_1}|\psi^{+}\rangle_{A_2} \pm |\psi^{-}\rangle_{A_1}|\psi^{-}\rangle_{A_2}\right)$ and $|\varphi\rangle_{12}^{\pm} = \frac{1}{\sqrt{2}}(|H\rangle_1|H\rangle_2 \pm |V\rangle_1|V\rangle_2)$ are

the Bell states. If the photons 1 and 2 are found to be in $|\varphi\rangle_{12}^+$ state, the SWs in the $A_1$ and $A_2$ modes will be projected onto the entangled state:

$$|\Phi\rangle_{A_1 A_2}^+ = \frac{1}{\sqrt{2}}\left(|\psi^+\rangle_{A_1}|\psi^+\rangle_{A_2} + |\psi^-\rangle_{A_1}|\psi^-\rangle_{A_2}\right).$$

The state $|\varphi\rangle_{12}^+$ can be identified by a suitable joint polarization measurement on the photons 1 and 2 [SR2].

For calculating the success probability of the conditionally prepared entangled state $|\Phi^+\rangle_{A_1 A_2}$, we have to consider the double excitations in either the spin-wave mode $A_1$ or $A_2$. In this case, the atom-photon entangled state is rewritten as:

$$|\Phi\rangle_{S_i - A_i} = \chi\left(\hat{S}_+^\dagger \hat{a}_L^\dagger + \hat{S}_-^\dagger \hat{a}_R^\dagger\right)|0\rangle_{S_i}^A |0\rangle_{A_i}^P + \frac{\chi^2}{2}\hat{S}_+^\dagger \hat{S}_-^\dagger \hat{a}_L^\dagger \hat{a}_R^\dagger |0\rangle_{S_i}^A |0\rangle_{A_i}^P \tag{S15}$$

As shown in Fig. 1(e), the two Stokes photons are guided into the PBS1 for performing the two-photon interference. If the two photons exit from two different output ports of PBS1 and enter the modes 1 and 2, the four-particle are in the state.

$$\begin{aligned}|\Psi\rangle_{G4} &= \frac{\chi^2}{2}\left(|\psi^+\rangle_{A_1}|\psi^+\rangle_{A_2}|H\rangle_1|H\rangle_2 + |\psi^-\rangle_{A_1}|\psi^-\rangle_{A_2}|V\rangle_1|V\rangle_2\right) \\ &+ \frac{\chi^2}{2}\left(|\psi^+\rangle_{A_1}|\psi^-\rangle_{A_1}|H\rangle_1|V\rangle_2 + |\psi^+\rangle_{A_2}|\psi^-\rangle_{A_2}|V\rangle_1|H\rangle_2\right)\end{aligned} \tag{S16}$$

Thus, conditioned on detecting a coincidence event between detectors $D_{H1}D_{H2}$ or $D_{V1}D_{V2}$ (see Fig. 1(e)), the two SWs will be projected into the state:

$$|\Psi\rangle_{G4} = \frac{\chi^2}{2}\left(|\psi^+\rangle_{A_1}|\psi^+\rangle_{A_2} + |\psi^-\rangle_{A_1}|\psi^-\rangle_{A_2}\right) + \frac{\chi^2}{2}\left(|\psi^+\rangle_{A_1}|\psi^-\rangle_{A_1} - |\psi^+\rangle_{A_2}|\psi^-\rangle_{A_2}\right) \tag{S17}$$

The first part is the required entangled states $|\Phi^+\rangle_{A_1 A_2}$, while the second part is the unnecessary two-excitation state. Since the probabilities of the two parts equal to each other, so the success probability of the conditionally prepared entangled state $|\Phi^+\rangle_{A_1A_2}$ is 1/2.

**Reference**

SR1. D. N. Matsukevich, T. Chanelière, M. Bhattacharya, S.-Y. Lan, S. D. Jenkins, T. A. B. Kennedy, and A. Kuzmich, Phys. Rev. Lett. **95,** 040405 (2005).

SR2. J.-W. Pan and A. Zeilinger, Phys. Rev. A. 57, 2208 (1998).